\documentclass[a4paper,11pt]{article}
\usepackage{pos}
\usepackage{wrapfig}
\usepackage{epsfig}

\usepackage{wasysym}
\newcommand{\postscript}[2]{\setlength{\epsfxsize}{#2\hsize}
   \centerline{\epsfbox{#1}}}

\definecolor{orange}{cmyk}{0,0.5,1,0}
\definecolor{rossoCP3}{cmyk}{0,.88,.77,.40}
\definecolor{graa}{rgb}{0.8,0.8,0.8}
\definecolor{blaa}{rgb}{0.2,0.2,0.6}

\title{Health threat from cosmic radiation during  manned missions to Mars}

\ShortTitle{Health threat from cosmic radiation during a manned mission to Mars}

\author*[a]{Alexandra D. Bloshenko}
\author[a]{Jasmin M. Robinson}
\author[b]{Rafael A. Colon}
\author[a]{Luis~A.~Anchordoqui}

\affiliation[a]{Department of Physics \& Astronomy,  Lehman College, CUNY, NY 10468, USA}
\affiliation[b]{Weill Cornell Medicine, Cornell University, NY 10065, USA}

\emailAdd{sasha.bloshenko@gmail.com}

\abstract{Cosmic radiation is a critical factor for astronauts' safety
  in the context of evaluating the prospect of future space
  exploration. The Radiation Assessment Detector (RAD) on board the
  Curiosity Rover launched by the Mars Scientific Laboratory 
  mission collected valuable data to model the energetic particle
  radiation environment inside a spacecraft during travel from Earth
  to Mars, and is currently doing the same on the surface of Mars
  itself. The Martian Radiation Experiment (MARIE) on board the Mars
  Odyssey satellite provides estimates of the absorbed radiation dose
  in the Martian orbit, which are predicted to be similar to the
  radiation dose on Mars' surface. In combination, these data provide
  a reliable assessment of the radiation hazards for a manned mission
  to Mars. Using data from RAD and MARIE we reexamine the risks for a crew on
  a manned flight to Mars and discuss recent developments in space exploration.}

\FullConference{37$^{\rm{th}}$ International Cosmic Ray Conference (ICRC 2021)\\
		July 12th -- 23rd, 2021\\
		Online -- Berlin, Germany}

\begin{document}
\maketitle

\section{Introduction}

Space -- the final frontier. It has almost become widely accepted that
it is our destiny to explore strange new worlds, to seek out new life
and new civilizations, to boldly go where no one has gone
before. However, humanity will have to find new ways to adapt to the
cold harsh environments of outer space if it is to ever become a
space-faring civilization. 

One of the greatest threats to long term
human survival during spaceflight is the exposure to high energy
radiation. There are
two primary types of radiation that humans are exposed to in space: solar energetic particle (SEP) events and Galactic cosmic
rays (GCRs)~\cite{Jakel}. Of the two, GCRs are a chronic source of high energy radiation that may damage the DNA of humans exposed to them for long periods of time, whereas the levels of damage caused by SEPs may vary drastically with the strength, duration, and location of the solar event~\cite{Barthel}.  The GCR flux fills the interplanetary space and is made up of
about 85\% hydrogen nuclei, 14\% helium nuclei, and around 1\%
high-energy and highly charged ($Z>2$) nuclei referred to as HZE
particles~\cite{Zyla:2020zbs}. Though there are fewer HZE particles than there are hydrogen
and helium nuclei, they possess significantly higher ionizing power,
greater penetration power, and a greater potential for the
radiation-induced damage to DNA. Therefore, determining the absorbed
radiation dosage of GCRs  and SEPs that an astronaut would receive during a
manned mission to Mars is necessary prior to any such mission. In this
paper we reexamine the radiation sources and risks for a crew on a manned flight to Mars.

One important factor of our study is
the duration of the trip to Mars. There are short duration (roughly
600 days with 30-day Mars stay time) or long
stay time (about 550 days with 180-day transfer to Mars and 180-day
inbound transfer) missions~\cite{Landau}. The usual benchmark here is the so-called Hohmann transfer, which is an elliptical orbit used to
transfer between the two circular planet orbits
using the lowest possible amount of propellant. The
launch from Earth for a successful Hohmann transfer must occur when Earth is
at perihelion and the landing takes place when  Mars is at 
aphelion. When launched within the proper window, a spacecraft will
reach Mars' orbit just as the planet moves to the same
point. Astronauts would travel from Earth to Mars, wait until the next
synchronized alignment between the planets (about 460 days), and
start the return trip to Earth. Kepler's 3rd law,
\begin{equation}
{\rm (orbital \ period/{\rm yr}})^2 = {\rm (Hohmann \ orbit \ semi\text{-}major \
  axis/{\rm AU}})^3 \,,
\end{equation}
provides a way to estimate the time-scale for the trip to Mars. Since
the mean Earth-Sun distance is $\overline D_{\odot-\oplus} = 1~{\rm AU}$
and the mean Mars-Sun distance is $\overline D_{\odot-\female} \simeq 1.542~{\rm AU}$, the Hohmann orbit
semi-major axis $= (\overline D_{\odot - \oplus} + \overline  D_{\odot - \female})/2 \simeq
1.262~{\rm AU}$. Therefore, the 
interplanetary travel will take 259~days each way, yielding a total mission
time of 978~days. We will use this time-scale to
estimate the radiation hazards for a manned mission to Mars. 

Before proceeding, we pause to specify and establish the relationship between the four different
units used to measure the amount of radiation absorbed by an object or
person, known as the ``absorbed dose.'' Such an absorbed dosed
reflects the amount of energy that radioactive sources deposit in
materials through which they pass.  The radiation absorbed dose (rad)
is the amount of energy (from any type of ionizing radiation)
deposited in any medium (e.g., water, tissue, air). An absorbed dose
of 1 rad means that 1 gram of material absorbed 100~erg of energy as a
result of exposure to radiation. The related international system unit
is the gray (Gy), where 1 Gy is equivalent to 100 rad.  Another
relevant unit is the Sievert (Sv), which is an ionizing radiation dose that measures the amount of energy absorbed by a
human body per unit mass (J/kg). This biological unit relates to the
previous units according to: 1~Sv = 1~Gy = 100~rad. Finally, the fourth important
unit to define is the Roentgen equivalent man (rem), which represents
the dosage in rads that will cause the same amount of biological
injury as one rad of X-rays or $\gamma$-rays: 1~Sv = 100~rem.

\section{Radiation hazards on space missions}
\label{sec:2}

\begin{wrapfigure}{r}{0.4\textwidth}
  \begin{center}
   \includegraphics[width=0.4\textwidth]{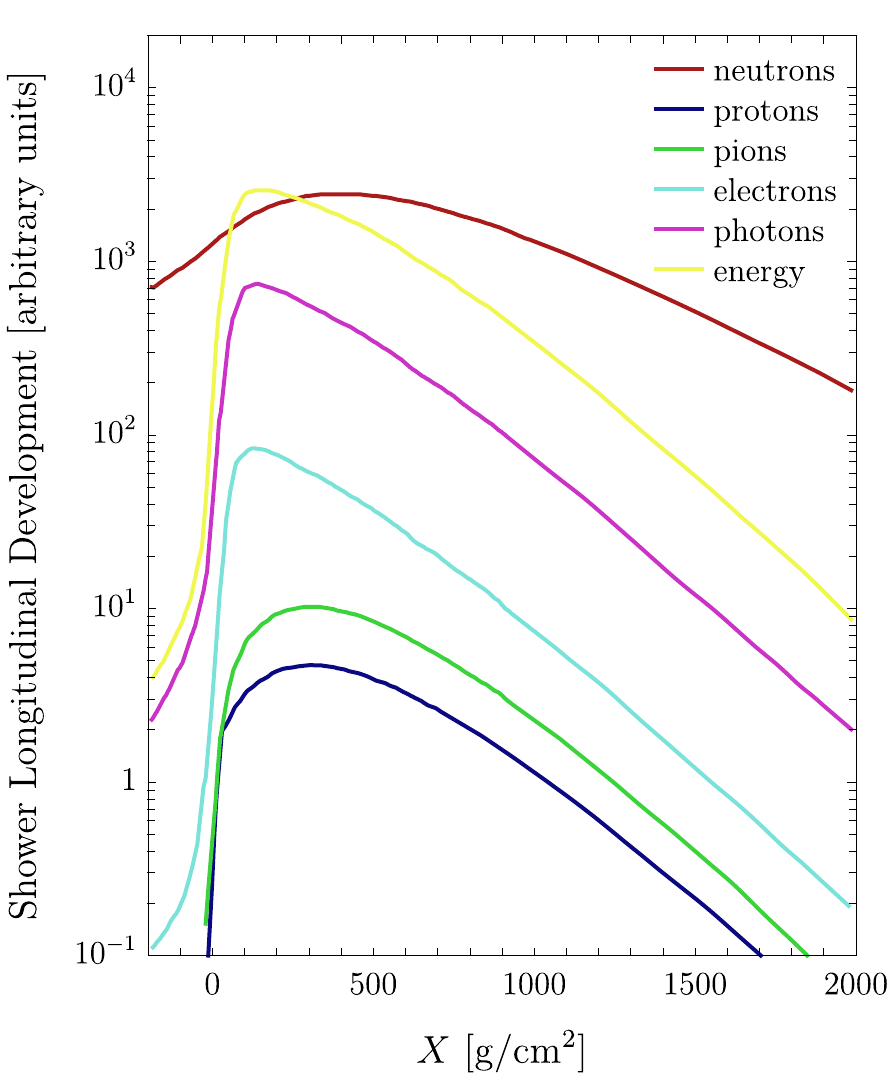}
  \end{center}
  \caption{Longitudinal development of a 100~GeV proton
    shower on lead as a function of depth $X$.
    \label{fig:1}}
\end{wrapfigure}

High-energy ($\gtrsim~{\rm GeV}$) protons and nuclei penetrating the spacecraft will
be subjected to different atomic and nuclear processes. These
relativistic particles 
will suffer inelastic collisions that produce, on average, a number of
fast secondary hadrons. Some of these hadrons (protons, neutrons,
charged pions) will have further nuclear collisions, resulting in a
hadronic cascade. Neutral pions, however, will decay almost instantly
and their decay products ($\gamma$-pairs) can initiate electromagnetic
showers, which are sustained by: {\it (i)}~electron-positron pair production by
photons; {\it (ii)}~Bremsstrahlung losses of electrons and positrons. As the
particle energy decreases, other processes become dominant (Compton
scattering of photons, photoelectric effect, ionization losses of
charged particles). Some of the charged pions and other hadrons will
also decay and produce muons. Besides the production of fast hadrons,
nuclear collisions also generate lower-energy (MeV) neutrons,
protons, light ions (alpha particles) and gamma rays, which are
emitted during the de-excitation of target nuclei. The protons and
light ions will mainly range out because of ionizing losses. The
neutrons, however, can be very penetrating and fully capable of
depositing a damaging dose deep inside tissue. 
As an illustration of all these processes, in Fig.~\ref{fig:1} we show the
longitudinal development of a simulated 100~GeV proton shower on lead,
using the program
FLUKA.

The overwhelming health risk comes from GCRs  since they exist as an isotropic background and are, therefore, consistently dangerous throughout the entire mission rather than at random intervals whenever SEP events occur. The energetic cosmic particles would strike the spacecraft to initiate hadronic cascades.  The
thicker the material they pass through, the more
secondary particles there will be in the cascade.  In condensed matter (liquid or solid)
the shower of a 1~GeV proton continues to grow until it reaches about $200~{\rm
  g/cm}^2$, which is the equivalent of about 2 meters of water.  After
this point the shower begins to attenuate.  The attenuation is
exponential, 
with a $1/e$ attenuation length of about $200~{\rm g/cm^2}$. The
attenuation length in the
Earth's atmosphere is somewhat smaller, roughly $140~{\rm
  g/cm}^2$.  In a gas, like in the Earth's atmosphere, the dependency
of the average depth of the peak
particle intensity $X_{\rm max}$ on energy can be parametrized as
$\langle X_{\rm max} \rangle = a + b \log_{10} (E/{\rm GeV})$, where $E$ is
the energy of the primary particle~\cite{Anchordoqui:2018qom}. For simulated $\gamma$-ray
showers, $a = 98~{\rm g/cm}^2$ and $b = 83~{\rm g/cm}^{2}$, whereas
for proton primaries $a = 111~{\rm g/cm}^2$ and $b = 74~{\rm
  g/cm}^2$~\cite{Schoorlemmer:2019cid}.
Whether pions decay or initiate more interactions accounts for the
attenuation difference between condensed matter and gas.

In the outer heliosphere, beyond about 100~AU, the slowing of solar
wind is thought to form a large magnetic barrier that shields out
$\gtrsim 90\%$ of the GCR radiation present in interstellar space at
energies below roughly 100~MeV~\cite{Florinski}. Because this reduction is so
large, even a very small change in the shielding efficiency can have a
large impact on the radiation environment in the solar
system. However, because these regions have never been directly
sampled or observed, there is great uncertainty about the physics of
outer heliospheric shielding and its sensitivity to changes in the
solar wind output and the local interstellar medium. A small fraction
of GCRs penetrate into the heliosphere and propagate toward the Sun
and planets. These residual GCRs are modulated by the solar wind's
magnetic field in the inner
heliosphere~\cite{Parker,Gleeson:1968zza}. The GCR intensity is at its
lowest during the peak of solar activity.  This is because the
enhanced solar activity sweeps away the low energy part of the GCR
spectrum.  The effect is statistically significant and more than compensates for solar generated cosmic rays during solar maximum.

Focusing now directly on Mars,  the red planet
 lost its magnetosphere 4 billion years ago and thus the solar wind
 interacts directly with the martian ionosphere, lowering the
 atmospheric density by stripping away atoms from the outer layer. The
 atmosphere of Mars consists of about 95\% carbon dioxide, 3\%
 nitrogen, 1.6\% argon and contains traces of oxygen,  water, and methane~\cite{Formisano}.
 The scale height ($\equiv$ vertical distance over which the density
 and pressure fall by a factor of $1/e$) of the martian atmosphere is about 10.8~km, which is higher than Earth's (8.5~km); the surface gravity of Mars is only about 38\% of Earth's, an effect offset by both the lower temperature and 50\% higher average molecular weight of the atmosphere of Mars.  The atmospheric pressure on the surface today ranges from a low of
0.030~kPa on Olympus Mons to over 1.155~kPa in
Hellas Planitia, with a mean pressure at the surface level of
0.60~kPa (which is only 0.6\% of that of the Earth 101.3~kPa). The highest atmospheric density on Mars is equal to the
density found 35~km above the Earth's surface. However, the atmosphere of Mars is just thick enough to produce plenty of
dangerous secondaries (including neutrons) from GCRs and SEPs. The radiation would be reduced to Earth-like
intensities for facilities located at a depth of
about $2,000~{\rm g/cm^2}$ beneath the martian surface.  

\begin{figure}[tpb]
\begin{minipage}[t]{0.49\textwidth}
\postscript{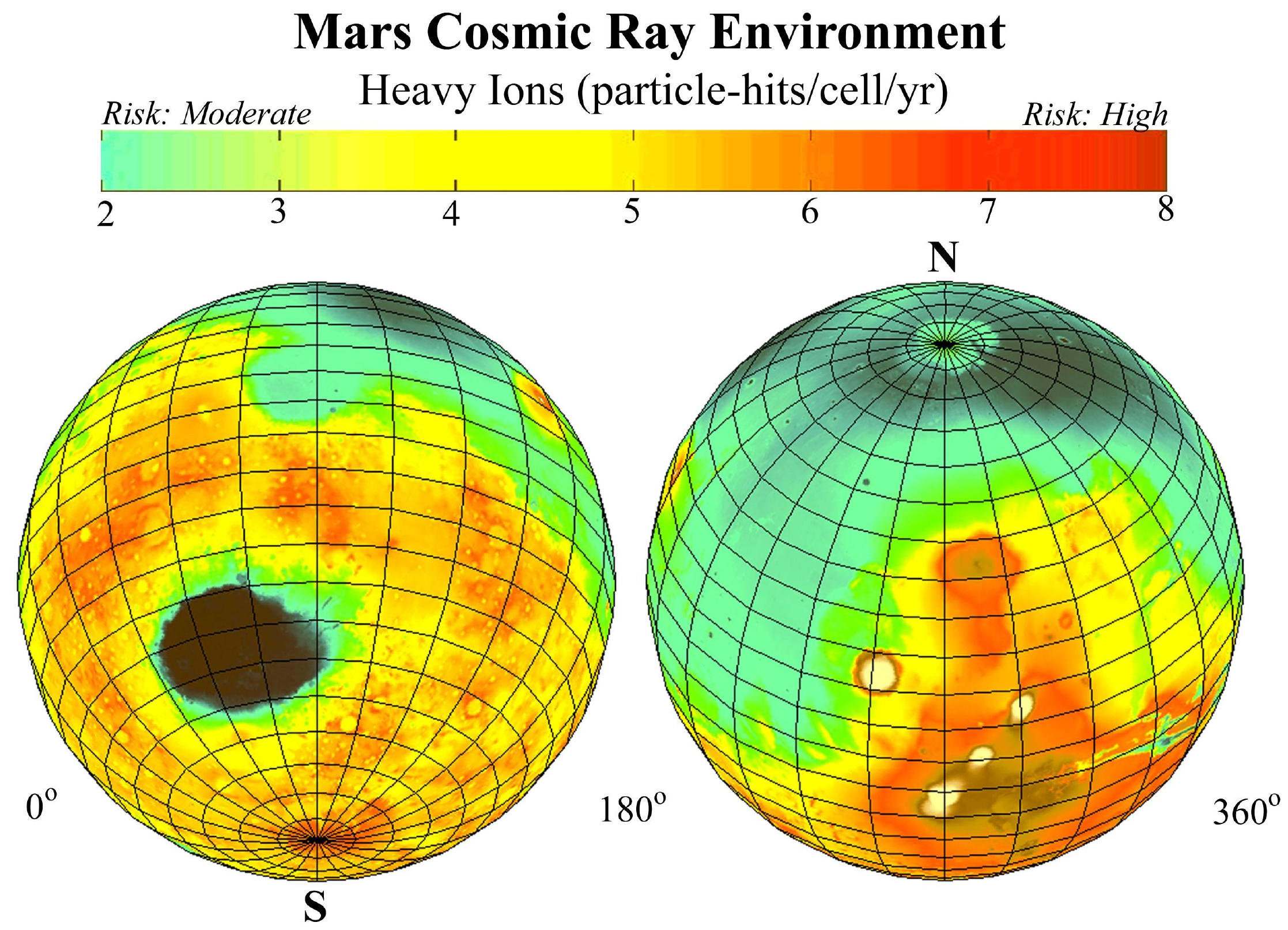}{0.90}
\end{minipage}
\begin{minipage}[t]{0.49\textwidth}
\postscript{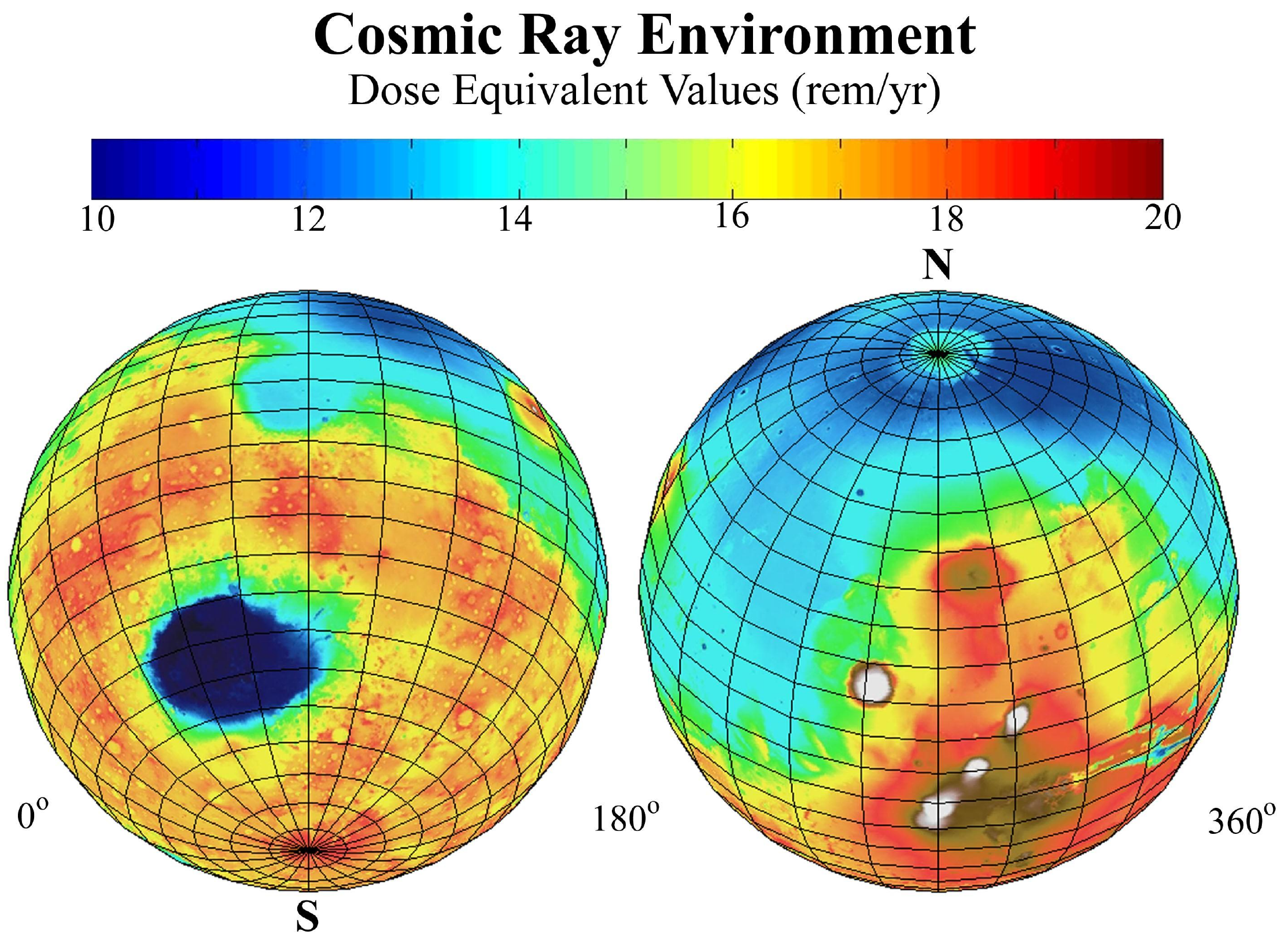}{0.90}
\end{minipage}
\caption{Estimated radiation dose from GCRs on
  the martian surface, from MARIE orbital radiation data and MOLA
  laser altimetry. The lower the altitude, the lower the expected
  dose, because the atmosphere provides shielding. The global map of
  Mars on the left panel shows estimates for amounts of high-energy
  particle cosmic radiation reaching the surface of the planet. Colors
  in the map refer to the estimated average number of times per year
  each cell nucleus in a human there would be hit by a high-energy
  cosmic ray particle. The range is generally from two hits, a
  moderate risk level, to eight hits, a high risk level. The global
  map of Mars on the right panel shows the estimated radiation dose
  from GCRs reaching the surface. The colors in the
  map refer to the estimated annual dose equivalent in rems. The range
  varies from 10~rem/yr to 20~rem/yr. This figure is
  courtesy of NASA/JPL/Johnson Space Center. \label{fig:2}}
\end{figure}

In our analysis we use data from the Radiation Assessment Detector
(RAD) on board Curiosity Rover launched by the Mars Scientific
Laboratory (MSL) mission~\cite{Zeitlin,Hassler,Guo} and the Martian Radiation Environment
Experiment (MARIE) on board the Mars Odyssey satellite~\cite{Zeitlin:2}. The RAD data
collection instrument gathered valuable data to model the energetic
particle radiation environment inside a spacecraft during travel from
Earth to Mars~\cite{Zeitlin,Guo}, and is currently doing the same on the surface of Mars
itself~\cite{Hassler}. The MARIE data collection instrument provides estimates of the
absorbed radiation dose in the martian orbit, which are predicted to
be similar to the absorbed radiation dose on the surface of
Mars~\cite{Zeitlin:2}. Variations of the GCR absorbed radiation dose
rate during the transfer between Earth and Mars range from 1.75 to
3.0~mSv/day~\cite{Zeitlin,Guo}. The GCR absorbed radiation dose on the surface of Mars measured by the
RAD instrument is 0.21 mGy/day~\cite{Hassler}.  The GCR absorbed radiation dose
in the Mar's atmosphere (collected by the MARIE instrument) is 0.25
mGy/day~\cite{Zeitlin:2}. In Fig.~\ref{fig:2} we show the
estimated radiation dose from GCRs on the
martian surface, using data from MARIE and the  Mars Orbiter Laser
Altimeter (MOLA) (which was one of five instruments on the Mars Global
Surveyor spacecraft that operated in Mars orbit from September 1997 to
November 2006~\cite{Smith}). The areas of Mars that are expected to
have the lowest levels of cosmic radiation are at the lowest points of
elevation, where there is more atmosphere above them, and at regions with
a localized magnetic field.

All in all, using these measurements we estimate that the absorbed radiation dose
during a round trip to Mars for a Hohmann transfer would be
between $906~{\rm mSv}$ and $1,554~{\rm mSv}$, whereas the absorbed
radiation dose during the stay on Mars would be between $97~{\rm mSv}$ and $115~{\rm mSv}$.

\section{Astronaut career radiation dose limits and prospects for space exploration}

Health risks from exposure to radiation on Earth in the form of 
X-ray or $\gamma$-ray radiation are well known. The
primary concern is the development of cancer due to radiation induced
mutations in DNA. DNA strand and tissue degradation, immunological
changes, cataracts, and damage to the central nervous system are all
serious risks that must also be mitigated in preparation for a manned
mission to Mars. These health risks are likely amplified in space,
where astronauts will face exposure to much higher energy radiation in
the form of GCRs and SEPs.

To protect astronauts from the dangerous effects of high-energy
radiation, the NASA space-program has determined a set of career radiation exposure limits for astronauts. Career radiation exposure limits vary by age and are lower for younger astronauts. The logic behind this is that younger astronauts have a longer life to live and therefore a greater chance of developing subsequent health problems in their lifetime. Career radiation exposure limits also vary by sex and are lower for female astronauts. This is due to a number of factors, including a generally higher risk of radiation induced cancer in females.

The NCRP Report No. 98 (NCRP 1989) made recommendations pertaining to astronauts involved in space shuttle missions and the
orbiting space station.  The longer follow-up of the atomic bomb survivors
and the evaluation of survivor organ doses by the DS86 dosimetry
system, used to recommend dose to risk conversion factors, led to
substantial increases in risk estimates~\cite{Preston:1987}.  The NCRP Report No. 132 (NCRP 2000) made new recommendations for radiation protection of
astronauts to accommodate the results of~\cite{Preston:1987}. The
predictions of increased risk factors in the NCRP 2000 report
are also supported by Mir space
station microdosimetry measurements~\cite{Cucinotta:2000}. However, a few years
after the publication of the NCRP Report No. 132, a reevaluation of
atomic bomb survivor organ doses with longer follow-up of the cohort
led to further increases in risk estimates~\cite{Preston:2004}. In the
last decade, NASA has deviated from the ground-based approach in which
radiation weighting factors are adopted to assign equal risk quality
factors for all types of cancer. In addition, the NCRP recommendation
for use of sex and age at exposure-dependent limits was extended to
consider a small specialized group of the population: a never-smoker
model, which lowers radiation risk estimates by about 20\% compared to
estimates for a U.S. average population. The NSCR-2012 model yields another increment in risk estimates,
with  substantially weaker dependence
on the age factor~\cite{Cucinotta:2013}. In
Table~\ref{tabla} we show a comparison between
the effective doses in the recent NSCR-2012 model and the values from
NCRP Reports No. 98 and No. 132~\cite{Cucinotta:2015}. (GCR solar modulation gives a $\sim
5\%$ effect~\cite{Cucinotta:2015}.) 

Setting side by side
the most recent risk factors given in Table~\ref{tabla} and the
absorbed dose estimates of Sec.~\ref{sec:2} it is straightforward to
see that (with current shielding technology) the Hohmann transfer is not
suitable for a round trip to Mars. However, a long stay mission of
about 550 days, with 900-day mission duration~\cite{Landau} would be
feasible. Indeed, this
is also consistent with the recent NASA twins
multidimensional analysis, which allowed comparison of the impact of the
spaceflight environment on one twin's human
health onboard  the
International Space Station for 340 days (his twin served as a
genetically matched ground control) to the simultaneous impact of the 
Earth environment~\cite{Garrett-Bakelman}. The data suggest that human health can be sustained over a year-long spaceflight.

\begin{table}[htb]
\vspace*{-0.1in}
\caption{Comparison of effective dose limits over 10~yr period. The last row corresponds
to never-smokers.}
\label{tabla}
\begin{tabular}{c|@{}cccc|@{}cccc}
\hline \hline
Sex & \multicolumn{4}{@{}c|}{Female} & \multicolumn{4}{@{}c}{Male} \\
\hline
Age (yr) & 25 & 35 & 45 & 55 &  25 & 35 & 45 & 55  \\
\hline
NCRP 1989 &~$1.00$~Sv & $1.75$~Sv & $2.50$~Sv  & $3.00$~Sv  &~$1.50$~Sv   & $2.50$~Sv  & $3.25$~Sv  & $4.00$~Sv \\
NCRP 2000 &~0.40~Sv  & 0.52~Sv & 0.75~Sv & 1.35~Sv  &~0.75~Sv& 1.00~Sv & 1.48~Sv
                                   & 2.98~Sv\\ 
NSCR 2012 & $-$ & 0.48~Sv & 0.51~Sv & 0.59~Sv & $-$ &  0.70~Sv & 0.75~Sv & 0.81~Sv  \\
NSCR 2012 & $-$ & 0.70~Sv & 0.75~Sv & 0.85~Sv & $-$ & 0.79~Sv & 0.85~Sv & 1.15~Sv\\
  \hline
  \hline
\end{tabular}
\end{table}

Given the perceived future of human spaceflight, there is a pressing
need to build upon the understanding of protecting astronauts from
HZE radiation. Heavier elements used in spacecraft shields actually produce
more secondaries than lighter elements, such as carbon and
hydrogen. This strongly supports the idea of looking into materials
made out of lightweight elements. One attractive material is the lightweight polyethylene
plastic called RFX1, which is 50\% more effective at shielding solar
flares and 15\% more effective at shielding GCR radiation than
aluminum is. However, while polyethylene plastic could improve the shielding, it
is not strong enough for load bearing aerospace structural
applications, which is why aluminum is primarily used. A new promising
material
under research at NASA  is made out of boron,
nitrogen, and hydrogen, and is called Boron Nitride Nanotubes (BNNT)~\cite{Thibeault}. Hydrogenated BNNT takes the form of microtubules - an excellent
design for increased stability, mechanical strength, and resistance to
extreme temperatures - and can also hold a large quantity of hydrogen
atoms for even greater shielding. Hydrogenated BNNT microtubules
can be woven into composites, fabric, yarn, and film forms that could
be integrated into the spacecraft structure as well as the astronauts'
spacesuits. While water is another molecule with a high hydrogen
content and potential to absorb radiation, it is significantly heavier
than polyethylene plastic and BNNT, and does not possess the necessary
strength for structural applications. Utilizing water
would require additional energy due to its mass and hence would be
more cost intensive. Creating an electrostatic radiation shield around
the spacecraft could also potentially deflect some cosmic radiation, but
would still leave astronauts exposed to dangerous levels of
radiation. 

In future missions to Mars, longer/permanent stays on the planet would
become a reality.  Protective measures on the red planet could include
constructing an underground shelter to reduce the amount of GCR
radiation that penetrates into human tissues. Such an underground
shelter would have to be built at $\sim 2,000~{\rm g/cm}^2$ depth for
effective shielding of secondaries produced by the interaction of HZE
particles with the atmosphere of Mars. There are several prospective
lava tubes on Mars that are located in low lying regions of the planet
and may be viable options for an underground shelter. As previously mentioned, lower elevations are locations with the least amount of cosmic
radiation and the hollow lava tubes may extend well below the surface
of Mars. Radiation experiments at analog lava tubes on Earth showed
that, on average, the amount of radiation in the interior of the tubes
is 82\% lower than on the surface~\cite{Paris:2020}. Given that Mars
is a differentiated terrestrial planet that formed from analogous
chondritic materials and that many of the same magmatic processes that
took place on Earth happened on Mars as well, it is possible that the
interior of lava tubes on Mars could also provide similar shielding
from radiation. For example, the radiation level at the Hellas Planitia  ($\sim
342~\mu$Sv/day) is considerably less than in other regions on the surface
of Mars ($\sim 547~\mu$Sv/day). A radiation exposure of
$342~\mu$Sv/day 
is, however, significantly higher than what humans are annually exposed to on Earth (levels typically range from
about 1.5 to 3.5~mSv/yr, but can be up to about 50~mSv/yr in some regions of
Europe).  Three candidate lava tubes identified in the vicinity of Hadriacus Mons (an ancient low-relief volcanic mountain
along the northeastern edge of Hellas Planitia)
could provide protection from excessive
radiation exposure. The absorbed dose inside these natural caverns would be $\sim 61.64~\mu$Sv/day.

\section{Conclusions}

We have shown that with current technology sending a
long stay time manned mission to
Mars, with 180-day transfer to Mars and 180-day inbound transfer, could be feasible. Incorporating BNNT shielding into spacecrafts
in conjunction with building underground shelter structures in lava
tubes on Mars might be the key for colonization of the red planet. Further considerations should include investments in space weather architecture to detect the onset of SEP events and provide a sufficient warning for astronauts to seek additional protection/shelter.

\section*{Acknowledgments}

This work has been supported by the U.S. National Science Foundation Grant PHY-2112257.


\begin{thebibliography}{99}

\bibitem{Jakel}
  O. J\"akel,
Z. Med. Phys. {\bf 14}, 267 (2004).  

\bibitem{Barthel}
J. Barthel and N. Sarigul-Klijn, Acta Astronautica {\bf 144}, 254 (2018).

\bibitem{Zyla:2020zbs}
P.~A.~Zyla \textit{et al.} [Particle Data Group],
PTEP \textbf{2020},  083C01 (2020).


\bibitem{Landau}
  D. F. Landau and J. M. Longuski,
 J. Spacecraft and Rockets
{\bf 43}, 1036 (2006). 




\bibitem{Anchordoqui:2018qom}
L.~A.~Anchordoqui,
Phys. Rept. \textbf{801}, 1 (2019).

\bibitem{Schoorlemmer:2019cid}
H.~Schoorlemmer, J.~Hinton and R.~L\'opez-Coto,
Eur. Phys. J. C \textbf{79},  427 (2019).

\bibitem{Florinski}
V. Florinski, G. P. Zank, and N. V. Pogorelov,
J. Geophys. Res. {\bf 108}, 1228 (2003).

\bibitem{Parker}
  E. N. Parker,
Planet. Space Sci. {\bf 13}, 9 (1965).


\bibitem{Gleeson:1968zza}
L.~J.~Gleeson and W.~I.~Axford,
Astrophys. J. \textbf{154}, 1011 (1968).


\bibitem{Formisano}
  V. Formisano, S. Atreya, T. Encrenaz, N. Ignatiev, and M. Giuranna,
 Science {\bf 306},
 1758 (2004).

\bibitem{Zeitlin}
  C. Zeitlin {\it et al.},
    Science {\bf 340}, 1080 (2013).

  
\bibitem{Guo}
  J. Guo {\it et al.},
 Astron. Astrophys. {\bf 577}, A58 (2015).


\bibitem{Hassler}
  D. M. Hassler {\it et al.},
 Science {\bf 343}, 1244797 (2014).

\bibitem{Zeitlin:2}
  C. Zeitlin {\it et al.},
   Adv. Space Res. {\bf 33}, 2204 (2004).

 \bibitem{Smith}
   D. E. Smith {\it et al.},
Science {\bf 284}, 1495 (1999).

\bibitem{Preston:1987}
  D. L. Preston and D. L. Pierce,
Hiroshima, Japan:
Radiation Effects Research Foundation; Report 9-87  (1987).


\bibitem{Cucinotta:2000}
F. A. Cucinotta, J. W. Wilson, J. R. Williams, and J. F. Dicello,  
Radiat. Meas.  {\bf 132}, 181 (2000).

\bibitem{Preston:2004}
D. L. Preston, D. A. Pierce, Y. Shimizu, H. M. Cullings, S. Fujita,
S. Funamoto, and K. Kodama,
  Radiat. Res. {\bf
    162}, 377  (2004).


\bibitem{Cucinotta:2013}
  F. A. Cucinotta, M. Y. Kim, and L. Chappell,
 Washington, DC: National Aeronautics and Space Administration; NASA TP 2013-217375 (2013).


\bibitem{Cucinotta:2015}
  F. A. Cucinotta, 
 Health Phys. {\bf 108}, 131 (2015).


\bibitem{Garrett-Bakelman}
  F. E. Garrett-Bakelman,
 Science {\bf 364}, eaau8650 (2019).
  
\bibitem{Thibeault}
S. A. Thibeault, C. C. Fay, G. Sauti,
J. H. Kang, and C. Park,
Patent Application Publication US 2015/0248941 A1, (2015). 

\bibitem{Paris:2020}
  A. J. Paris, E. T. Davis, L. Tognetti, and C. Zahniser,
J. Wash. Acad. Sci. (in press)
[arXiv:2004.13156].

\end{thebibliography}
\end{document}